# Decentralized and Coordinated V-f Control for Islanded Microgrids Considering DER Inadequacy and Demand Control

Buxin She, *Student Member, IEEE,* Fangxing (Fran) Li, *Fellow, IEEE,* Hantao Cui, *Senior Member, IEEE,*
Jinning Wang, *Student Member, IEEE,* Liang Min, *Senior Member, IEEE,*
Oroghene Oboreh-Snapps, *Student Member, IEEE,* Rui Bo, *Senior Member, IEEE*

*Abstract*— **This paper proposes a decentralized and coordinated voltage and frequency (V-f) control framework for islanded microgrids, with full consideration of the limited capacity of distributed energy resources (DERs) and V-f dependent load. First, the concept of DER inadequacy is illustrated with the challenges it poses. Then, a decentralized and coordinated control framework is proposed to regulate the output of inverter-based generations and reallocate limited DER capacity for V-f control. The control framework is composed of a power regulator and a V-f regulator, which generates the supplementary signals for the primary controller. The power regulator regulates the output of grid-forming inverters according to the real-time capacity constraints of DERs, while the V-f regulator improves the V-f deviation by leveraging the load sensitivity to V-f. Next, the static feasibility and small signal stability of the proposed method are rigorously proven through mathematical formulation and eigenvalue analysis. Finally, a MATLAB-Simulink simulation demonstrates the functionalities of the control framework. A few goals are fulfilled within the decentralized and coordinated framework, such as making the best use of limited DERs' capacity, enhancing the DC side stability of inverter-based generations, and reducing involuntary load shedding.**

## I. INTRODUCTION

A typical microgrid is composed of multiple distributed energy resources (DERs), energy storage systems, and local loads, which can operate in either grid-connected mode or islanded mode [1]. Compared with a conventional bulk power system, microgrids have the characteristics of more DERs, smaller system size [2], higher uncertainty [3]-[4], lower system inertia [5]-[6], and stronger voltage and frequency (V-f) coupling [7]-[8]. All these features create challenges for load-generation balance and V-f regulation in microgrids, especially in islanded microgrids that are not supported by the main grid.

A hierarchical control framework, which consists of the primary controller, secondary controller, and tertiary controller, has been widely used in islanded microgrids [9]. The primary

B. She, F. Li, and J. Wang are with the Dept. of EECS, The University of Tennessee, Knoxville, TN, USA.
H. Cui is with the School of ECE, Oklahoma State Univ., Stillwater, OK, USA.
M. Liang is with the Precourt Institute for Energy, Stanford University, Stanford, California, USA.
O. Oboreh-Snapps and R. Bo are with Dept. of ECE, Missouri Univ. of Science and Technology, Rolla, MO, USA.

controller has the highest bandwidth and is responsible for automatic load sharing. Due to its fast response and important power-sharing functionality, the primary control dominantly determines the V-f deviation and stability of microgrids [10]. Inductive microgrids usually employ the *P-f* and *Q-V* droop control in primary control, while resistive microgrids use the reverse *P-V* and *Q-f* droop curves [11]-[12]. In this way, V-f are maintained after normal disturbances.

An islanded microgrid has many inverter-based generations. As a result of the high penetration of renewable energy, a major cause of grid instability and large V-f deviation pertains to the inadequacy of DERs. DER inadequacy indicates that the capacity of inverter-based generations is insufficient to supply a microgrid's total load. The impacts of limited DER capacity can be seen at both the inverter level and the grid level. For a single inverter, the DC side voltage maintained by the buck-boost converter tends to ripple first and then droop across the linked capacitor when DERs cannot supply enough power [13]. For the whole microgrid, insufficient active and reactive power (P-Q) may result in over frequency and voltage dip [14]. The negative impacts of insufficient DERs necessitate more efficient use of existing generation capacity, which makes the power regulation of inverter-based generations critical in islanded microgrids. The output of grid-following (GFL) inverters can be controlled easily by changing the $P_{ref}$ and $Q_{ref}$. Grid-forming (GFM) inverters, on the other hand, make it hard to control power accurately and quickly. GFM inverters are controlled as voltage sources, and their output is mostly determined by the grid side performance.

Some research has been focused on limiting the output of GFM inverters. [15]-[16] designed a current limiter and put it between the current and voltage regulator. The threshold of the current limiter is usually set at 2 p.u. to 3 p.u. because it is designed to prevent high currents in abnormal conditions, e.g., the high currents in low voltage ride through after grounded faults [17]. Apart from the current limiter, [13] and [18] added a virtual impedance to the voltage control loop to limit the terminal voltage of the inverter. However, the current limiter and virtual impedance are mainly designed for protecting the hardware device and cannot accurately control the output of inverters in normal conditions. Moreover, the saturation caused by the current limiter and virtual impedance may result in the instability of the microgrids [19]. In fact, the output of GFM inverters in islanded microgrids is determined by the load-sharing results. Their output can be regulated by changing the power-sharing parameters, such as nominal $P_{ref}$-$Q_{ref}$ and droop gains. The droop gains are tuned proportionally to DER capacities, allowing each DER to take on new loads according to its capacity [20]. Considering that a DER's capacity may fluctuate from time to time due to its intermittent nature, [21]-[25] developed some adaptive droop strategies, but each with different control targets. [21]-[22] aimed at better reactive power-sharing. [23] focused on better transient performance. [24] presented a method for suppressing circuiting current between inverters. [25] addressed inverter-based load control. Although the adaptive control algorithms improve the load sharing results, they are no longer effective if the overall DER capacity is insufficient. This is caused by the saturation of various control modules. Thus, a new GFM inverter power regulator considering DER inadequacy has yet to be developed. Further, since P-Q generation shares the same apparent capacity, it is critical to enable the coordination between P-Q generation while regulating the output of inverters [26].

Based on the above discussions, this paper developed a decentralized and coordinated V-f control framework to address DER inadequacy in islanded microgrids. With the proposed control methods, each GFM inverter can output any power according to real-time capacity constraints without communicating with adjacent inverters. In addition, demand control [27]-[28] is incorporated into this control framework to reduce load-generation mismatch when DER capacity is insufficient. Demand control leverages load sensitivity to V-f and minimizes involuntary load shedding. Mathematical analysis and simulation results show that demand control collaborates well with inverter output regulator within the decentralized and coordinated framework. The contributions of this paper are as follows.

- We summarize the challenges posed by DER inadequacy in islanded microgrids and address them with full consideration of DER capacity constraints, demand control, and the coordination between P-Q and V-f control.
- We propose a decentralized and coordinated V-f control framework. The control framework can regulate the output of each GFM inverter accurately and quickly. It also improves V-f deviation and reduces involuntary load shedding without depending on knowledge and measurements from adjacent nodes.
- We mathematically prove the static feasibility and small signal stability of the control framework and demonstrate the proposed method using MATLAB-Simulink in a multi-inverter islanded microgrid.

The rest of this paper is organized as follows. Section II discusses the concept and challenges of DER inadequacy in islanded microgrids, and then introduces the foundation of power and V-f regulation. The decentralized and coordinated V-f control framework is proposed in Section III, followed by theoretical proofs of static feasibility and small signal stability in Section IV. Section V demonstrates the control method with a modified Banshee microgrid [29]. Conclusions are drawn in Section VI.

## II. DER Inadequacy in Islanded Microgrids

This section presents the challenges related to DER inadequacy in islanded microgrids and introduces the foundation of the control framework to address these challenges.

### A. Configuration of islanded microgrids

#### 1) Grid-forming inverter

A droop-controlled GFM inverter is the basic unit of islanded microgrids. Fig. 1 shows the diagram of a GFM inverter supplying static V-f dependent load. The GFM inverter is controlled as a voltage source, and the control framework consists of a current regulator, a voltage regulator, and a primary regulator, as shown in (1)-(3). As a negative feedback controller, the purpose of the primary regulator below is to emulate the droop characteristics of conventional synchronous generators and automatically share the load among microgrid inverters.

$$\begin{bmatrix} e_d \\ e_q \end{bmatrix} = \begin{bmatrix} v_d \\ v_q \end{bmatrix} + wL_f \begin{bmatrix} -i_q \\ i_d \end{bmatrix}$$
$$+ \begin{bmatrix} k_{pid} + \dfrac{k_{iid}}{s} & 0 \\ 0 & k_{piq} + \dfrac{k_{iiq}}{s} \end{bmatrix} (\begin{bmatrix} i_{dref} \\ i_{qref} \end{bmatrix} - \begin{bmatrix} i_d \\ i_q \end{bmatrix}) \quad (1)$$

$$\begin{bmatrix} i_{dref} \\ i_{qref} \end{bmatrix} = \begin{bmatrix} i_{gd} \\ i_{gq} \end{bmatrix} + wC_f \begin{bmatrix} -v_q \\ v_d \end{bmatrix}$$
$$+ \begin{bmatrix} k_{pud} + \dfrac{k_{iud}}{s} & 0 \\ 0 & k_{puq} + \dfrac{k_{iuq}}{s} \end{bmatrix} (\begin{bmatrix} v_{dref} \\ v_{qref} \end{bmatrix} - \begin{bmatrix} v_d \\ v_q \end{bmatrix}) \quad (2)$$

$$\begin{bmatrix} w_{ref} \\ V_{ref} \end{bmatrix} = \begin{bmatrix} k_{df} & 0 \\ 0 & k_{dv} \end{bmatrix} (\begin{bmatrix} P_0 \\ Q_0 \end{bmatrix} - \begin{bmatrix} P_m \\ Q_m \end{bmatrix}) + \begin{bmatrix} w_0 \\ V_0 \end{bmatrix} \quad (3)$$

Where $P_0$ and $Q_0$ are initial power setting points; $P_m$ and $Q_m$ are measured output; $k_{df}$ and $k_{dv}$ are frequency and voltage droop gains, respectively.

#### 2) V-f dependent load

As shown in (4), V-f dependent ZIP loads are modeled in this paper [30].

$$\begin{cases} P_l = P_0(p_1 V_l^2 + p_2 V_l + p_3) \begin{bmatrix} 1 + K_{pf}(f - f_0) \end{bmatrix} \\ Q_l = Q_0(q_1 V_l^2 + q_2 V_l + q_3) \begin{bmatrix} 1 + K_{qf}(f - f_0) \end{bmatrix} \end{cases} \quad (4)$$

Where $V_l$ is the load side voltage; $p_1 + p_2 + p_3 = 1$ and $q_1 + q_2 + q_3 = 1$; $K_{pf}$ is the sensitivity of $P_l$ to $f$; $K_{qf}$ is the sensitivity of $Q_l$ to $f$.

The coefficients in (4) represent the compositions of ZIP loads. They change with the time of day and week, seasons, and weather [31], but can be assumed as constant in the timescale of primary control. Because microgrids are usually small in scale, the change in inverter terminal voltage is quickly reflected at the load side, which in turn changes the system demand that depends on the load sensitivity to V-f [10]. As a result, demand control can participate in V-f regulation in islanded microgrids.

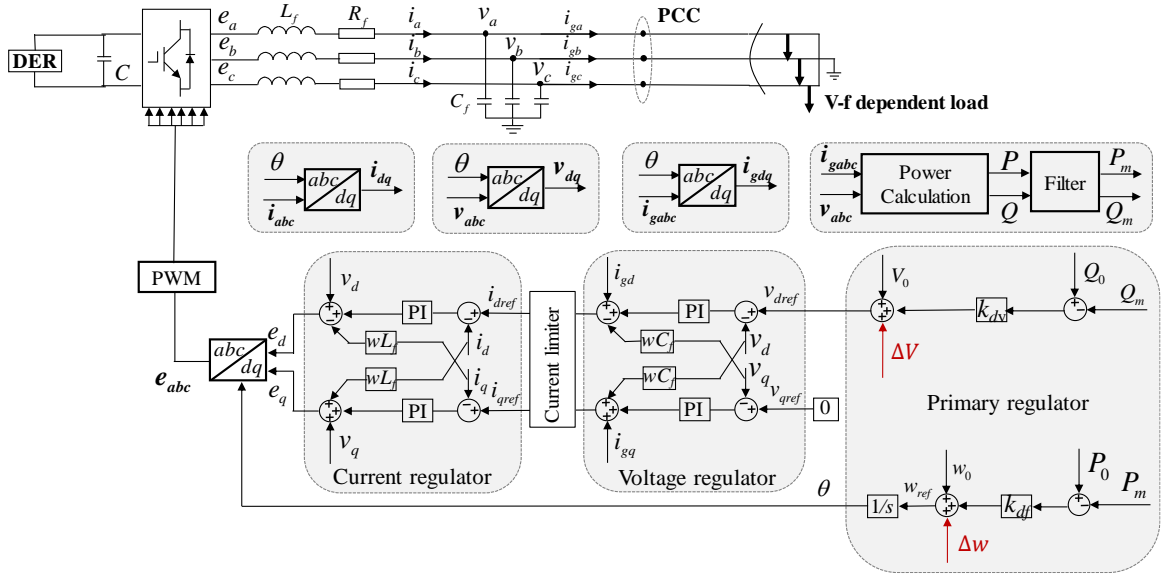

Fig. 1 Diagram of a droop-controlled GFM inverter supplying V-f dependent load.

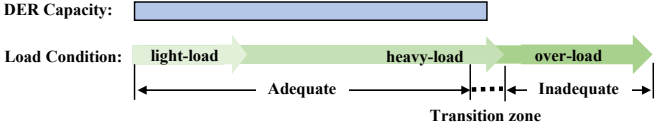

Fig. 2 DER inadequacy under various load level

### B. DER inadequacy and its challenges

An islanded microgrid is a self-sufficient system. DERs pose great challenges to grid operation due to their intrinsic features of uncertainty and intermittency.

#### 1) Load condition and DER inadequacy

Fig. 2 shows a diagram of DER inadequacy under different load levels. With the DER capacity as the baseline, there are generally three load conditions, i.e., light-load, heavy-load, and over-load. 1) In the light-load condition, the total load is much smaller than the DER capacity. Thus, grid V-f may not significantly deviate from the nominal value after normal load change. 2) In the heavy-load condition, the total load is close to, but slightly under, the DER capacity. The output of some GFM inverters may exactly equal the capacity of the DC side source. 3) In the over-load condition, the total load is larger than the DER capacity, such that the power balance cannot be guaranteed unless conducting load shedding.

In Fig. 2, if the total load is significantly lower than the generation capacity, the DER is adequate; if the total load is much larger than the generation capacity, the DER is inadequate. The load-generation balance in the critical transition area between the adequate and inadequate zones merits more attention. In this paper, we investigate this critical transition area for its potential to improve stability and reduce V-f deviation, which is critical in the background of high penetration of intermittent renewable generation and extreme weather conditions.

#### 2) Challenges

There are three main challenges in islanding microgrids when it comes to DER inadequacy, including improper load sharing, DC side instability, and large V-f deviation.

The first challenge is related to load sharing. In the conventional droop control, the load sharing results are determined by the initial setting point [$P_0$, $Q_0$] and droop gains [$k_{dp}$, $k_{dv}$]. The load change at an equilibrium point is shared proportionally with the droop gains and among the microgrid inverters, which are tuned based on the DER capacity. However, due to the intermittency of DERs, their capacities may fluctuate from time to time. The GFM inverter has a stability issue if load sharing cannot adapt to the real-time DER capacity [32]. The challenge is more critical in the transition zone shown in Fig. 2 because most GFM inverters work in critical states with little reserved capacity.

The second challenge stems from improper load sharing among DERs and is related to DC side voltage. When the shared loads exceed the capacity of the DC side DERs, there is insufficient energy to support the capacitor voltage, which may result in a voltage dip first and then distort Pulse-width modulation (PWM). Furthermore, the inverter-based generation will trip and compound the DER inadequacy. Fig. 3(a) shows an unexpected DC voltage dip and generation loss when the shared loads exceed the DER capacity at 5s.

The third challenge relates to large V-f deviation. The first two challenges often impact a single or several inverters, while the third challenge, which is about allocating limited generation capacity to the V-f regulation loop, impacts the whole grid. Fig. 3(b) shows the bounded generation constraint. Following the same capacity constraint, V-f regulation results are quite different when implementing distinct P-Q generation strategies. Under bounded generation constraints, a large $k_{df}$ means more active power decrease and more frequency dip, while a large $k_{dv}$ means more reactive power decrease and more voltage dip. Fig. 3(c) shows the unexpected V-f regulation with distinct droop gains. Usually, the deviation of frequency should be less than 0.01 p.u. and the deviation of voltage should be less than 0.05 p.u.. In Fig. 3(c), the requirements for V-f deviation cannot be met at the same time. V-f deviation can be improved according to the supplementary signals generated by the conventional secondary controller. However, due to saturation caused by the capacity constraints in Fig. 3(b), the inverter may fail to respond to these signals, resulting in an unexpected DC voltage dip.

Considering the challenges discussed above, it is urgent to develop a V-f control framework that can: 1) regulate the output of GFM inverters and improve load sharing results based on real-time DER capacity; 2) adjust P-Q generation under the condition and constrained DER capacity for both acceptable voltage and frequency deviation.

### C. Foundation of the decentralized and coordinated V-f control framework

In Fig. 1, the primary control signals $w_{ref}$ and $v_{dref}$ determine the load sharing results and V-f deviation. The foundation of the proposed control method is to design a power regulator and a V-f regulator, which generate the supplementary signals for $w_{ref}$ and $v_{dref}$ based on real-time measurements. Specifically, the supplementary signal of the power regulator is generated according to the error between real-time reference capacity $S_r$ and measured output $S_m$, which automatically re-shares the exceeding output among the other inverters. The supplementary signal of the power regulator guarantees the constrained generation shown in Fig. 3(b). Then, like the conventional secondary control, the V-f regulator can further adjust P-Q generation along the boundary and perform load power control through V-f regulation. Finally, the proposed control framework can regulate GMF inverter output and improve V-f deviation at the same time.

As in the conventional droop control, all the supplementary signals are generated depending on local measurements. The proposed control method further enables cooperation between

- *Generation and load*: Generation and load power are regulated simultaneously till reaching a new equilibrium.

- *V regulation and f regulation*: Requirements for V-f deviation are considered at the same time. A tradeoff is made while allocating limited DER capacity for V-f regulation.

- *P generation and Q generation*: Limited DER capacity is allocated to P-loop and Q-loop properly.

As a result, the proposed method is decentralized and coordinated. Section III dives into the control framework.

## III. PROPOSED DECENTRALIZED AND COORDINATED CONTROL FRAMEWORK

This section introduces the proposed decentralized and coordinated V-f control framework. As shown in Fig. 4(a), the control framework consists of a power regulator and a V-f regulator. They are implemented locally, generating supplementary signals $\Delta w'$ and $\Delta V'$ for the primary controller of each GFM inverter.

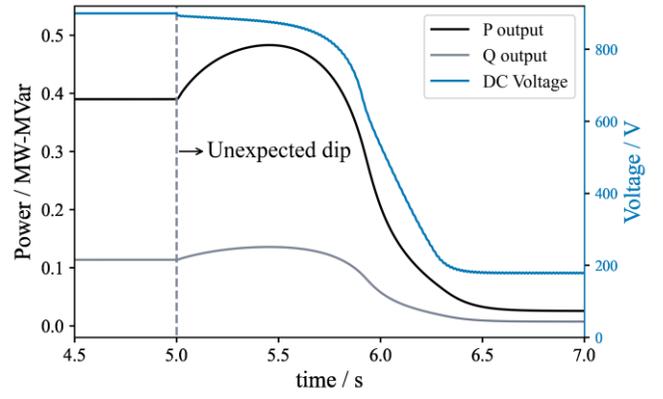

(a)

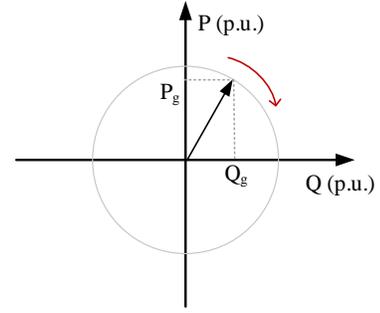

(b)

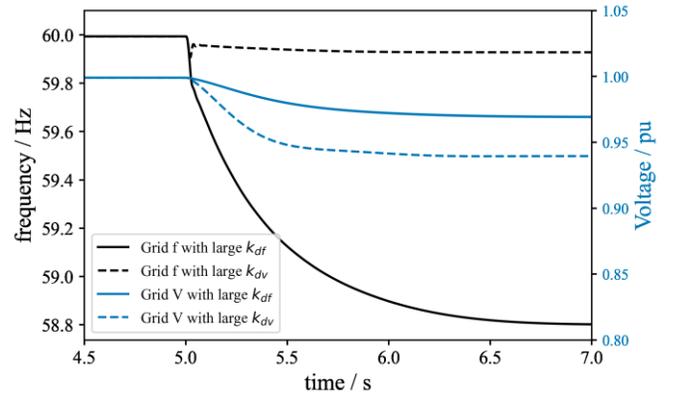

(c)

Fig. 3 Challenges related to DER inadequacy: (a) DC side voltage dip; (b) Constrained generation capacity for V-f regulation; (c) Unexpected V-f regulation with distinct droop gains.

### A. Power regulator

The transfer function of the power regulator is shown in (5)-(6).

$$\begin{bmatrix} \Delta w_1 \\ \Delta V_1 \end{bmatrix} = (k_{ps} + \frac{k_{is}}{s})e_s \begin{bmatrix} k_w \\ k_v \end{bmatrix} \tag{5}$$

$$e_s = \begin{cases} 0, & \text{if } S_r > S_m \\ S_r - S_m, & \text{if } S_r \leq S_m \end{cases} \tag{6}$$

The errors between $S_r$ and $S_m$ are fed to a proportional-integral (PI) controller, which is then allocated to the frequency and voltage regulation loop according to regulation gains $k_w$ and $k_v$, respectively. Typically, if the inverter outputs more power than $S_r$, $e_s$ is negative, and the inverter will decrease its output by decreasing the terminal voltage and angle frequency. The final decrease was determined by $e_s$ proportionally allocated to the active power loop and the reactive power loop, specifically, the proportion between $k_w$ and $k_v$. Given load sensitivity and disturbance, a larger $k_w$ means more sharing in the active power loop and more frequency dip, while a larger $k_v$ means more sharing in the active power loop and more frequency dip.

It is worth noting that $e_s$ allocation by the power regulator was not always optimized due to the uncertainties from both the generation and load sides. There are no single-value, fixed $k_w$ and $k_v$ that can handle all the disturbance scenarios at the same time. Hence, we proposed the V-f regulator to generate additional supplementary signals and further adjust the active power and reactive power sharing results. In practice, we can set $k_w$ and $k_v$ based on load power factor (PF), so as to match the load-side disturbance as much as possible. Specifically, if PF = $\cos\theta$, then set $k_w/k_v = \tan\theta$.

### B. V-f regulator

The transfer function of the V-f regulator is shown in (7)-(9).

$$\begin{bmatrix} \Delta w_2 \\ \Delta V_2 \end{bmatrix} = \begin{bmatrix} k_{pf} + \dfrac{k_{if}}{s} & 0 \\ 0 & k_{pv} + \dfrac{k_{iv}}{s} \end{bmatrix} \boldsymbol{Tr}\left(\begin{bmatrix} e_f \\ e_v \end{bmatrix}\right) \quad (7)$$

$$e_f = \begin{cases} \Delta f + \Delta f_{max}, & \text{if } \Delta f < -\Delta f_{max} \\ 0, & \text{if } -\Delta f_{max} \leq \Delta f \leq \Delta f_{max} \\ \Delta f - \Delta f_{max}, & \text{if } \Delta f > \Delta f_{max} \end{cases} \quad (8)$$

$$e_v = \begin{cases} \Delta V + \Delta V_{max}, & \text{if } \Delta V < -\Delta V_{max} \\ 0, & \text{if } -\Delta V_{max} \leq \Delta V \leq \Delta V_{max} \\ \Delta V - \Delta V_{max}, & \text{if } \Delta V > \Delta V_{max} \end{cases} \quad (9)$$

Where $\Delta f = f_0 - f_m$, $\Delta V = V_0 - V_m$, and $\boldsymbol{Tr}$ is the trigger logic shown in Fig. 4(b).

The diagram of the trigger logic is shown in Fig. 4(b), which considers the relationship between V-f regulation and load-generation balance. In step 1, it estimates the state of V-f and determines the required capacity for V-f regulation. For example, if $e_w > 0$ ($state_f = 1$), more reactive power is required for frequency regulation, and frequency regulation will take up more generation capacity; in reverse, $e_w < 0$ means less need of active power and frequency regulation could give up some generation capacity. This logic is also applicable to the reactive power loop (V regulation loop). Then in step 2, the trigger logic judges whether the existing DER capacity is adequate or not to meet the V-f deviation requirements, which is categorized into the following three conditions.

- "$state_f = state_v = 1$" means V-f droop too much at the same time, and both V-f regulation loops need more generation capacity. The existing DER capacity is not sufficient to regulate V-f, thus load shedding is needed.
- "$state_f = state_v = 0$" means V-f deviations are within the acceptable region and no additional regulation signal is needed.
- The remaining combinations of $state_f$ and $state_v$ mean that one regulation (V or f) loop needs more generation capacity and the other could give up some generation

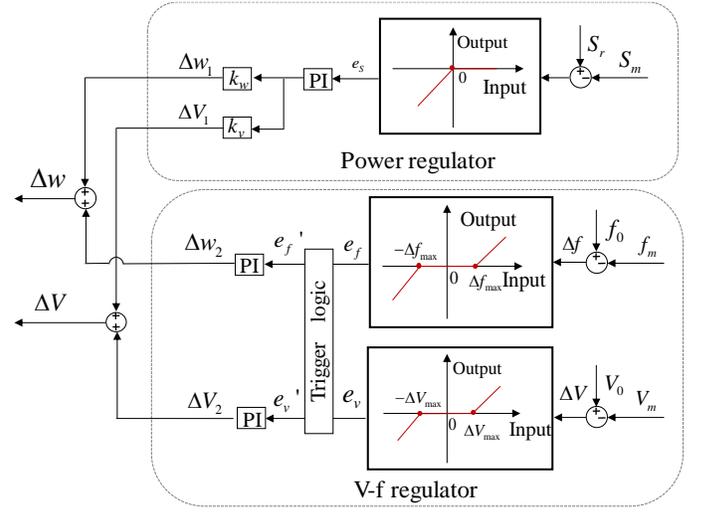

(a)

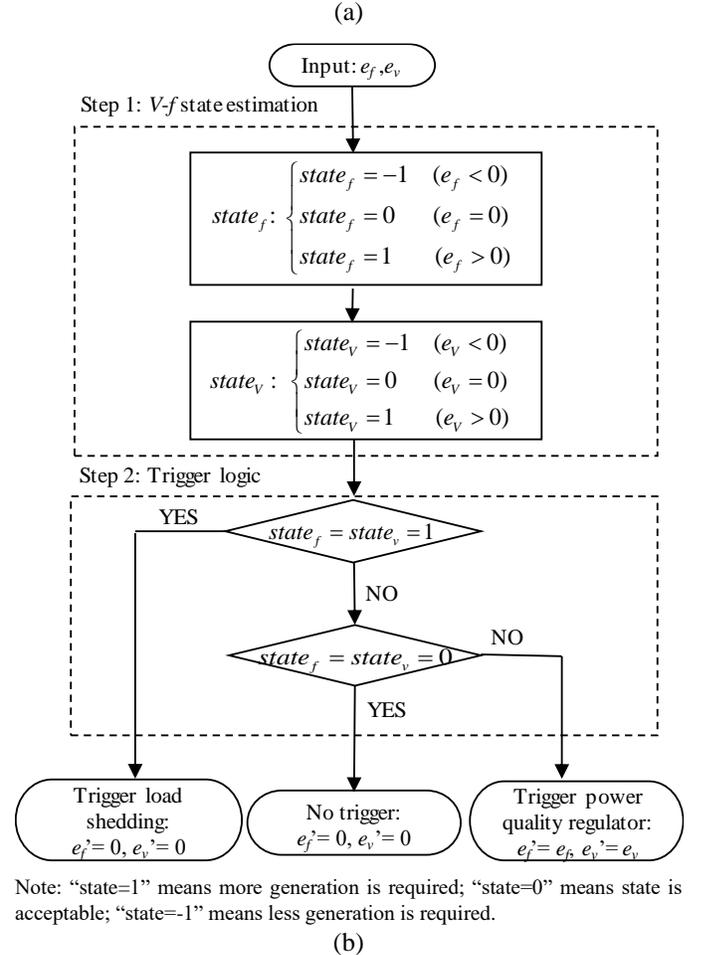

Note: "state=1" means more generation is required; "state=0" means state is acceptable; "state=-1" means less generation is required.

(b)

Fig. 4 Diagram of decentralized and coordinated control: (a) control framework (b) trigger logic of V-f regulator.

capacity. Then, grid V-f could be maintained without the need for load shedding.

Note that the power regulator has priority over the V-f regulator because it guarantees the capacity constraints that are closely related to DC voltage stability. Hence, a power regulator usually has larger controller gains and higher bandwidth than a V-f regulator.

## IV. Static Feasibility and Small Signal Stability

This section proves static fixability and small signal stability of the proposed control framework through algebraic formulation and eigenvalue analysis.

### A. Static feasibility

The proposed microgrid control framework has two main functionalities: 1) regulate the output of GFM inverters through load re-sharing; 2) improve V-f deviation by reallocating generation capability along the constrained boundary in Fig. 3(b). The mechanism of load re-sharing is similar to that of the adaptive droop gain method, which has been clearly illustrated in [21]. Hence, this subsection mainly verifies the second functionality through algebraic formulation.

The V-f regulator is identical to the conventional secondary controller when DER capacity is adequate, which improves V-f deviation by adjusting the output of inverter-based generations. It changes to a constrained secondary controller in the transition zone when the load is close to the total generation. In the following discussion, the static equilibrium is derived when the total generation is insufficient, based on which the state change in the transition zone is further derived.

#### 1) Static equilibrium

With the focus on a general islanded microgrid formed by $N$ inverters, each inverter is connected to an independent bus with a local V-f dependent load. Under droop control and the proposed framework, there are $6N$ independent equations, including $2N$ droop equations in (10), $2N$ load equations in (11), and $2N$ power flow equations in (12).

$$\begin{cases} f = f_{0,i} + k_{df}(P_{inv,i} - P_{inv0,i}) \\ V_i = V_{0,i} + k_{dv}(Q_{inv,i} - Q_{inv0,i}) \end{cases} \quad (10)$$

$$\begin{cases} P_{l,i} = P_{l0,i}(p_1 V_i^2 + p_2 V_i + p_3)\left[1 + K_{pf}(f - f_0)\right] \\ Q_{l,i} = Q_{l0,i}(q_1 V_i^2 + q_2 V_i + q_3)\left[1 + K_{pf}(f - f_0)\right] \end{cases} \quad (11)$$

$$\begin{cases} P_i = P_{inv,i} + P_{l,ii} = G_{ij}V_i^2 - G_{ij}\sum_{i\neq j}V_iV_j\cos\theta_{ij} - B_{ij}\sum_{i\neq j}V_iV_j\cos\theta_{ij} \\ Q_i = Q_{inv,i} + Q_{l,i} = G_{ij}V_i^2 - G_{ij}\sum_{i\neq j}V_iV_j\cos\theta_{ij} - B_{ij}\sum_{i\neq j}V_iV_j\cos\theta_{ij} \end{cases} \quad (12)$$

where $f$ is the global steady state frequency; and $f_{0,i}$ and $V_{0,i}$ are the nominal frequency and voltage of the $i$th inverter, respectively. Assume the reference inverter has a power angle $\theta$=0, and then there are $6N$ decision variables in (10)-(12), including 1 global frequency, $N$ voltage, $N$-1 power angle, $N$ active power output, $N$ active load, $N$ active inverter output, and $N$ reactive inverter output. The number of decision variables equals the total number of equations. Thus, the new equilibrium after a given disturbance can be calculated, allowing further estimation of the proposed method to reduce load shedding.

#### 2) State transition under constrained DER capacity

In initial equilibrium, $P_{inv,i}$=$P_{inv0,i}$ and $Q_{inv,i}$=$Q_{inv0,i}$. Then, assume a step increase in the base load and capacity constraints on the inverters.

$$\begin{cases} \text{Load}: \begin{cases} P_{l0,i}' = P_0 + \Delta P \\ Q_{l0,i}' = Q_0 + \Delta Q \end{cases} \quad \forall i = 1, 2, \cdots, N \\ \text{Generation}: P_{inv,i}'^2 + Q_{inv,i}'^2 = S_i^2 \end{cases} \quad (13)$$

The original equilibrium will transition to a new equilibrium while also meeting the network power flow and power balance constraints as follows.

$$\begin{cases} P_i' = P_{inv,i}' + P_{l,i}' \\ \quad = G_{ij}V_i'^2 - G_{ij}\sum_{i\neq j}V_i'V_j'\cos\theta_{ij}' - B_{ij}\sum_{i\neq j}V_i'V_j'\cos\theta_{ij}' \quad \forall i, j, i \neq j \\ Q_i' = Q_{inv,i}' + Q_{l,i}' \\ \quad = G_{ij}V_i'^2 - G_{ij}\sum_{i\neq j}V_i'V_j'\cos\theta_{ij}' - B_{ij}\sum_{i\neq j}V_i'V_j'\cos\theta_{ij}' \quad \forall i, j, i \neq j \end{cases} \quad (14)$$

where the prime symbols mark the new state variables.

During the transition, the conventional droop control is rendered ineffective, and the $2N$ droop equations are replaced by $N$ bounded capacity constraints. This gives us freedom to design the new generation output. Given that $P_{inv,i}'$ and $Q_{inv,i}'$ meet the capacity constraints in (13), there are $4N$ state variables and $4N$ equations left. Then, for each $(P_{inv,i}', Q_{inv,i}')$ point on the capacity circle, the corresponding new equilibrium V-f deviation is solvable, as the number of equations and decision variables are identical.

### B. Small signal stability

The power system model is composed of some differential equations and algebraic equations [33, Chapter 9]. The proposed power regulator and V-f regulator are incorporated into the conventional droop control controller in Fig. 1. Then, the complete dynamic model of microgrids is expressed as a set of differential-algebraic equations (DAEs) as follows.

$$\begin{cases} \dot{x} = f(x, y) \\ 0 = g(x, y) \end{cases} \quad (15)$$

where $x$ is the vector of state variables and $x$=[$\varphi_{vs}$, $\varphi_{fs}$, $\varphi_v$, $\varphi_f$, $\varphi_{vd}$, $\varphi_{vq}$, $\varphi_{id}$, $\varphi_{iq}$, $\varphi_p$, $v_d$, $v_q$, $i_d$, $i_q$, $P_m$, $Q_m$, $i_{gd}$, $i_{gq}$]; $y$ is the vector of algebraic variables and $y$ = [$v_{dref}$, $v_{qref}$, $i_{dref}$, $i_{qref}$, $e_d$, $e_q$, $P$, $Q$].

After linearizing the DAEs in (15), the small-signal model is obtained in (16). Then, the state matrix $A$ is calculated by eliminating the algebraic variables. Since the V-f regulator consists of piecewise functions, each linear part can be analyzed according to initial operating points.

$$\begin{bmatrix} \Delta \dot{x} \\ 0 \end{bmatrix} = \begin{bmatrix} f_x & f_y \\ g_x & g_y \end{bmatrix} \begin{bmatrix} \Delta x \\ \Delta y \end{bmatrix} \quad (16)$$

$$A = f_x - f_y g_y^{-1} g_x \quad (17)$$

Note that small signal stability can only guarantee the stability of the system around a specific equilibrium. As (8) and (9) are nonlinear functions with dead bands, time-domain simulation (TDS) was required to validate the stable transition between equilibriums, as detailed in Section V.

### C. Exemplified in an ideal system

This subsection exemplified the static feasibility and small signal stability of an ideal three-inverter system in Fig. 5.

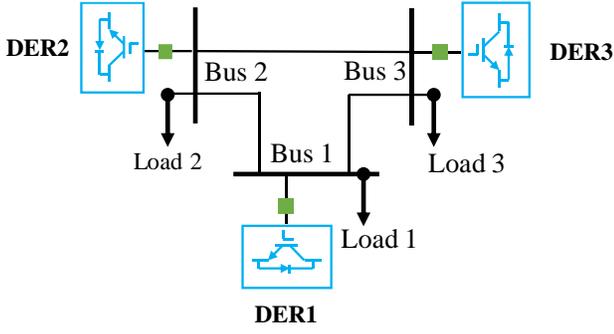

Fig. 5 Single-line diagram of the ideal three-inverter system.

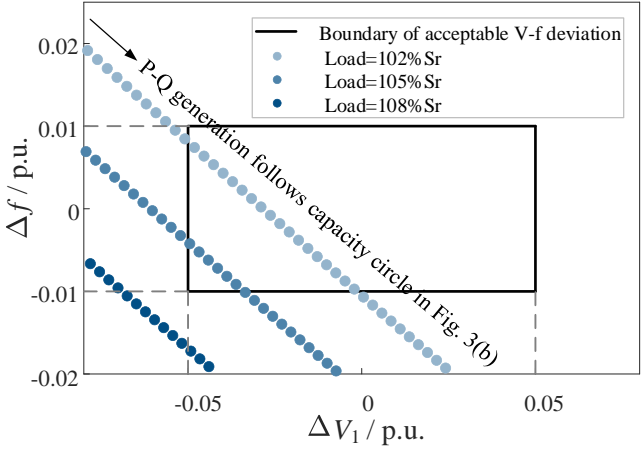

Fig. 6 V-f deviation under bounded generation constraints.

### 1) Static feasibility

Assume an intentional load increase at the initial operating point ($P_0$, $Q_0$). Then, (13) and (14) predict the new equilibrium after an intentional load change under the condition that the conventional primary controller becomes invalid. Fig. 6 shows the new equilibriums when implementing different generation strategies under various load increases. To facilitate observation, the high-dimensional state space is projected into 2-dimensional $\Delta V_1$-$\Delta f$ space, where $\Delta V_1$ is Bus 1 voltage deviation. Based on Fig. 6, we have the following observations:

a) For Load=102%$S_r$ and Load=105%$S_r$, although the total load is larger than the generation capacity, load-generation balance can still be achieved without performing load shedding because there are some new equilibriums inside the black security rectangle that meet the V-f deviation requirements.

b) If the current operating point is located outside the black security rectangle where the V-f regulator may be triggered, the P-Q generation can be adjusted along the bounded generation circle until the operating point comes back to the acceptable region.

c) If the total load is too large, e.g., Load=108%$S_r$, load shedding is required since there is no operating point inside the security rectangle. But the shed load could be less than conventional strategies to reduce unnecessary involuntary load shedding. For example, 3% to 5% is enough to improve V-f deviation using demand control.

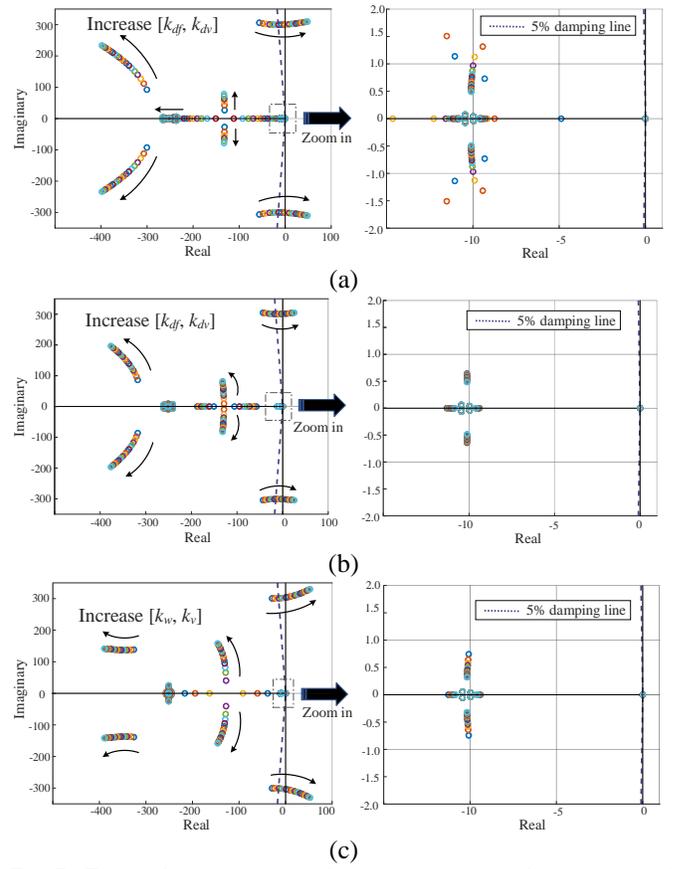

Fig. 7. Eigenvalue trajectories when sweeping control parameters (a) Condition 1: crossing droop gain [$k_{df}$, $k_{dv}$]=[0.025, 0.125]; (b) Condition 2: crossing droop gain [$k_{df}$, $k_{dv}$]=[0.0165, 0.0825]; (c) Condition 3: crossing power regulator gain [$k_w$, $k_v$]=[0.0825, 0.0619].

### 2) Small signal stability

We then choose an operating point outside the black security rectangle in Fig. 6 and calculate the eigenvalues of $A$. To verify the robustness of the proposed method, we plot the eigenvalues by sweeping control parameters in three different ways.

- Condition 1: sweep droop gain [$k_{df}$, $k_{dv}$] without the integration of the proposed control framework.

- Condition 2: sweep droop gain [$k_{df}$, $k_{dv}$] with the integration of the proposed control framework.

- Condition 3: sweep power regulator gain [$k_w$, $k_v$] with the integration of the proposed control framework.

The eigenvalue trajectories under Conditions 1, 2, and 3 are shown in Figs. 7(a), 7(b), and 7(c), respectively, where the critical gains that force the eigenvalue to cross the $y$ axis are also marked in the title. Based on Fig. 7, we have the following observations:

a) In Conditions 1 and 2, all the eigenvalues are on the left half-plane before and after the integration of the proposed method. Then, increasing the droop gains could result in instability, which shows that the system maintains a specific stability margin if tuned properly.

b) Condition 2 has smaller crossing gains than those of Condition 1. The integration of the proposed framework method could decrease the stability margin of the original system, but it still preserves a specific yet sufficient margin for increasing the power regulator gains, as shown in Condition 3.

## V. Case Study in A Real Microgrid

This section verifies the functionalities of the proposed method in a modified real microgrid, and further demonstrates its transient performance between different equilibriums.

### A. Case overview

The microgrid shown in Fig. 8 operates in islanded mode with switch 100 off. It is modified from the Banshee distribution system by replacing the diesel generator on BUS 103 with a battery energy storage system (BESS) and connecting a PV and a BESS to BUS 102 and 105, respectively [29]. The three GFM inverters are named G1, G2, and G3 in Fig. 7. The microgrid also supplies the static V-f dependent ZIP load and motor load [29]. The load parameters and inverter control parameters are listed in Tab. 1 and Tab. 2, respectively. All the simulations were run in MATLAB® version R2020a, with a PC Intel® Core i7-8665U CPU at 2.10 GHz and 16 GB RAM.

Two test scenarios are designed to validate the key functionalities of the proposed control framework. Scenario 1 focuses on inverter power regulation, while Scenario 2 focuses on the cooperation of the power regulator and V-f regulator, which leverage the limited DER capacity and demand control to improve V-f deviation and reduce involuntary load shedding. In addition, we choose current limiter as a baseline to show the advantages of the proposed method.

### B. Scenario 1: inverter power regulation

Following the timeline, the basic settings of Scenario 1 are three-fold:

a) Before 8 s, G1-G3 are controlled with the conventional droop method represented in Fig. 1.

b) At 8s, the inverter power regulators of G1-G3 are implemented and triggered immediately. Meanwhile, the real-time reference capacities of G1-G3 are set as 1.2MVA, 0.6MVA, and 2.0 MVA, respectively.

c) At 12s, load C2 connected to Bus 106 increases by 200kW, 100kVar.

#### 1) Performance of power regulator

Fig. 9 shows the microgrid performance in Scenario 1, where Fig. 9(a) is the dynamic output of G1-G3, Fig. 9(b) is the static output, and Fig. 9(c) is the dynamic voltage and frequency.

In Fig. 9(a), the output of G1 and G3 follows the left axis, while the output of G2 follows the right axis, based on which we have the following observations:

- According to the initial setting point ($P_0$, $Q_0$) and droop gains ($k_{df}$, $k_{dv}$) before 8 s, all the load is automatically shared among G1, G2, and G3 as 1.24MVA, 0.67MVA, and 1.24MVA, respectively.

- At 8s, S1 and S2 violate the real-time capacity constraints, and the power regulator starts working. Then, P1, Q1, P2, and Q2 decrease until S1 and S2 are regulated to the reference value.

- Even though there is a sudden load increase at 12 s, S1 and S2 still converge to the reference value. Meanwhile, P3, Q3, and S3 increase at 8 s and 12 s to compensate for the decreased generation and the increased load.

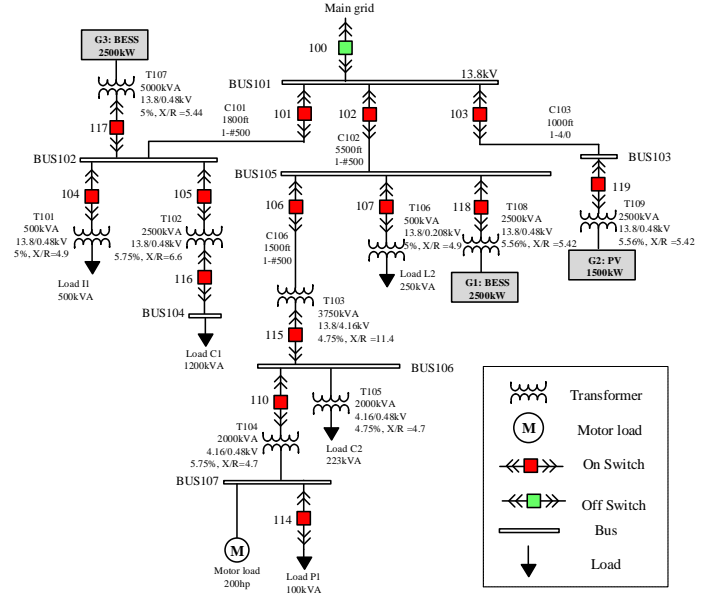

Fig. 8. Single line diagram of the islanded microgrid [29].

Table. 1 ZIP load parameters

| Load | $P$ composition/ [$P_1, P_2, P_3$] | $Q$ composition/ [$q_1, q_2, q_3$] | V-f sensitivity/ [$k_{pf}, k_{qf}$] |
|---|---|---|---|
| L1 | 0.1, 0.3, 0.6 | 0.5, 0.3, 0.2 | 2, -0.1 |
| L2 | 0.6, 0.2, 0.2 | 0.6, 0.2, 0.2 | 2.5, -2 |
| C1 | 0.5, 0.3, 0.2 | 0.2, 0.3, 0.5 | 3, -0.1 |
| C2 | 0.4, 0.3, 0.3 | 0.4, 0.3, 0.3 | 1, -0.5 |
| P2 | 0.5, 0.3, 0.2 | 0.5, 0.3, 0.2 | 3, -1 |

Table. 2 Inverter control parameters

| Parameter | | G1 | G2 | G3 |
|---|---|---|---|---|
| Filter | $L_f$/H | $5\times10^{-5}$ | $2.5\times10^{-5}$ | $5\times10^{-5}$ |
| | $C_f$/F | $1\times10^{-5}$ | $1\times10^{-5}$ | $1\times10^{-5}$ |
| Current regulator gains [$k_p, k_i$] | | [0.5, 2] | [0.5, 2] | [0.5, 2] |
| Voltage regulator gains [$k_p, k_i$] | | [0.1, 1] | [0.1, 1] | [0.1, 1] |
| Droop gains [$k_{df}, k_{dv}$] | | [0.01, 0.05] | [0.005, 0.025] | [0.01, 0.05] |
| Power regulator gains [$k_{ph}, k_{hs}, k_{ss}, k_i$] | | [0.2, 4, 0.008, 0.006] | [0.1, 2, 0.004, 0.003] | [0.2, 4, 0.008, 0.006] |
| V-f regulator gains [$k_{pf}, k_{if}, k_{pv}, k_{iv}$] | | [0.05, 1, 0.05, 1] | [0.05, 1, 0.05, 1] | [0.05, 10, 0.05, 1] |

Fig. 9(b) further presents three static operating points at 6s, 10s, and 14s. In Fig. 9(b), the circle and square operating points of G1 and G2 lie exactly on the capacity boundary after triggering the power regulator.

In Fig. 9(c), the voltage and frequency are measured at the terminal of G3, and they hardly change after the power regulator starts working. This indicates that the power regulator is able to regulate the output of GMF inverters without significantly affecting the grid's voltage and frequency.

#### 2) Comparison with the current limiter

The existing current limiter is mainly designed for protecting devices in abnormal conditions [15]-[16], and the threshold is usually set at 2 p.u. to 3 p.u.. To make it more comparable with our proposed method, we assume that the threshold is adaptively updated according to the real-time DER capacity. In addition, the DER capacity is preferentially allocated to the frequency regulation loop. Then, the performance of the current limiter is shown in Fig. 10. After triggering the current limiter, the output of G1 and G2 is limited to their real-time generation capacities. However, the sudden limit results in larger voltage and frequency deviations around 8 s, as detailed in Fig. 10(c).

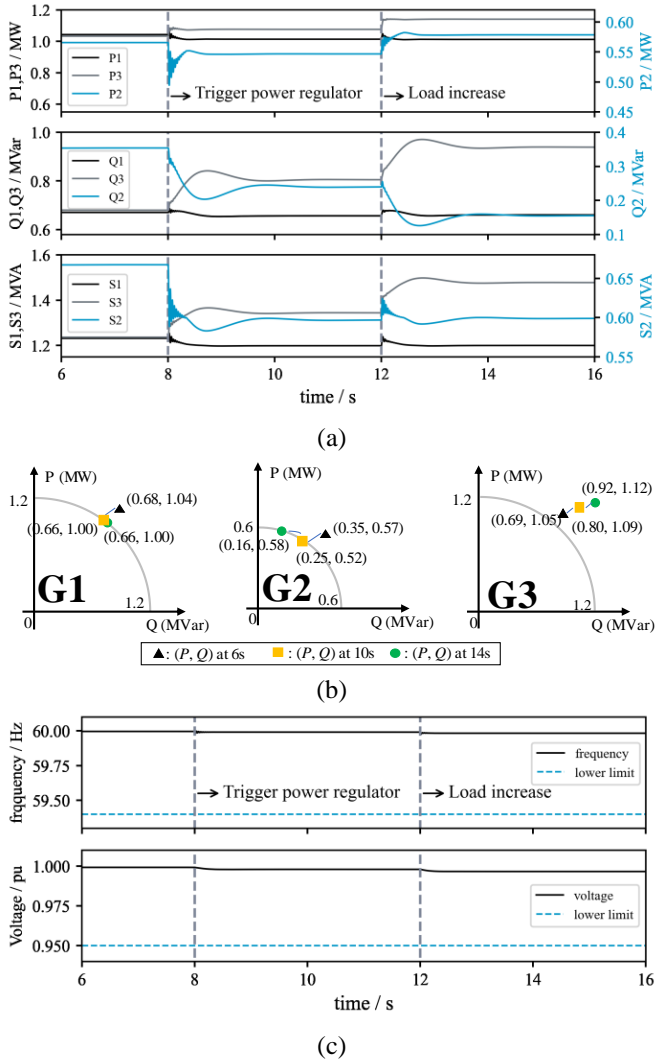

Fig. 9. Performance of power regulator in Scenario 1: (a) dynamic inverter output; (b) static inverter output; (c) dynamic frequency and voltage

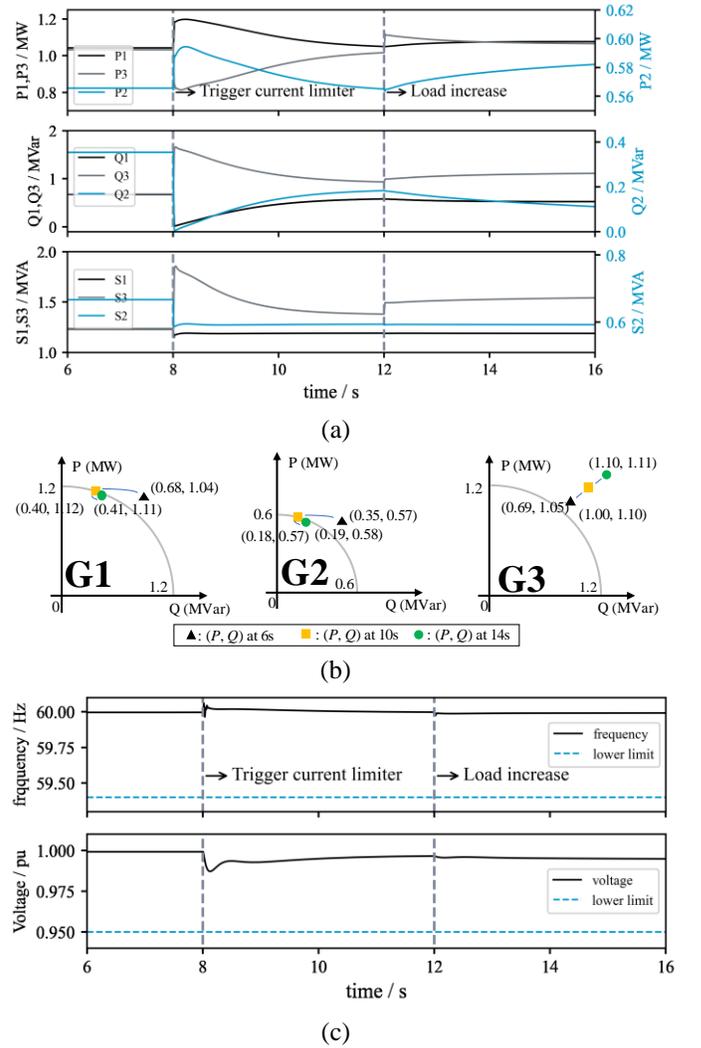

Fig. 10. Performance of current limiter in Scenario 1: (a) dynamic inverter output; (b) static inverter output; (c) dynamic frequency and voltage

### C. Scenario 2: cooperation of power regulator and V-f regulator

In the transition zone shown in Fig. 2, the conventional primary and secondary controllers become invalid due to the saturation caused by insufficient DER capacity. Then, the cooperation of power regulator and V-f regulator becomes important. Scenario 2 is further divided into two parts, Scenario 2-1 and Scenario 2-2, to show their cooperation under various load change. Specifically, load C2 increases by 100kW, 200kVar in Scenario 2-1 while by 200kW, 100kVar in Scenario 2-2. Following the timeline, Scenario 2-1 and Scenario 2-2 have the same settings as follows, and their grid performance is shown in Figs. 11-12.

a)  Before 8 s, G1-G3 are controlled with the conventional droop method represented in Fig. 1.

b)  At 8 s, load C2 increases.

c)  At 12 s, the reference capacities of G1-G3 are set as 1.2MVA, 0.6MVA, and 1.2MVA, respectively. The total load is thus larger than the total generation capacity, and the power regulator starts working.

d)  At 16 s, the V-f regulator starts working.

#### 1) Performance of the proposed control framework

Fig. 11 and Fig. 12 show the performance of proposed control method in Scenarios 2-1 and 2-2, respectively. They both demonstrate the cooperation of the power regulator and V-f regulator.

In Figs. 11(a) and 12(a), as a result of different load change, Scenario 2-1 and Scenario 2-2 have distinct load sharing results after 8 s. At 12 s, the power regulator starts working. S1, S2, and S3 are regulated to 1.2MVA, 0.6MVA, and 1.2 MVA around 12.5 s, which are exactly equal to the reference capacities. The reduced generation is compensated by the load

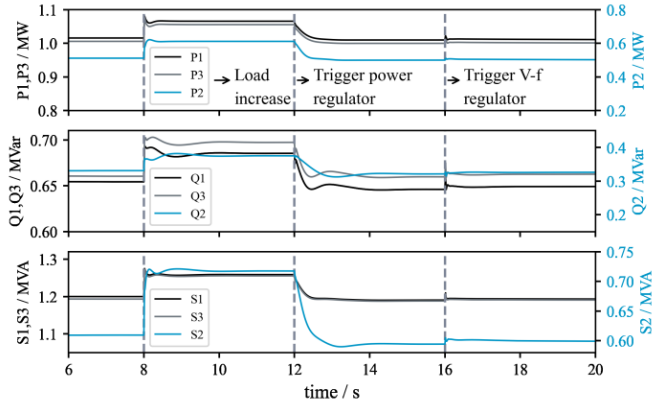

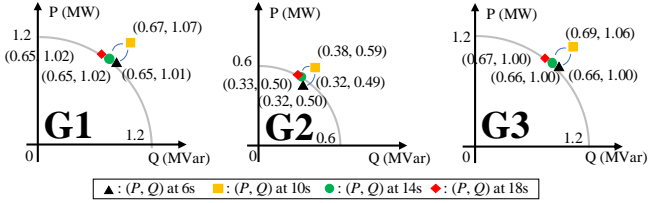

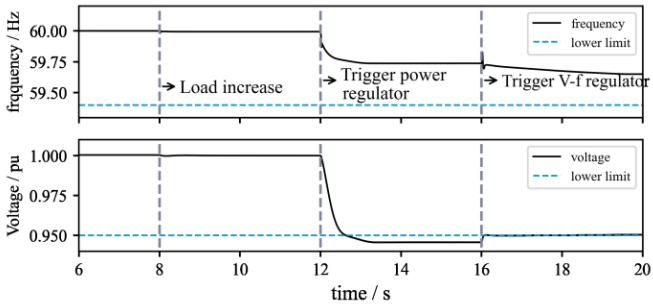

(a)

(b)

(c)

Fig. 11. Performance of the proposed control framework in Scenario 2-1: (a) dynamic inverter output; (b) static inverter output; (c) dynamic frequency and voltage

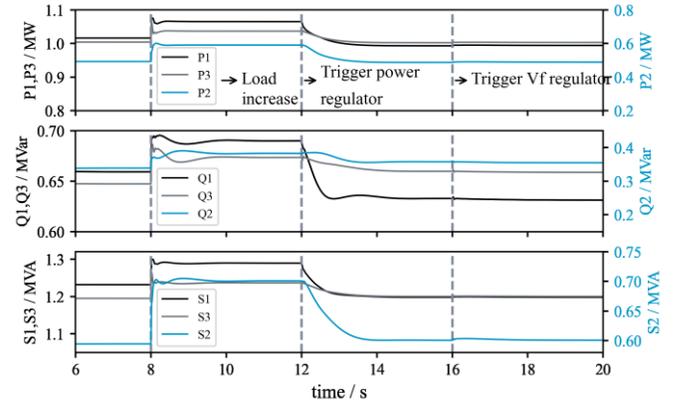

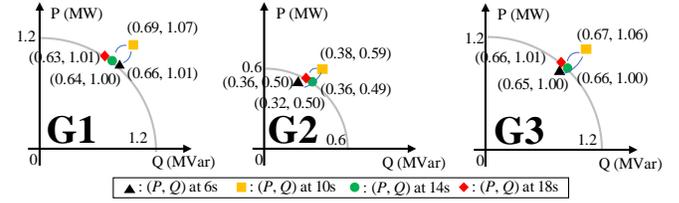

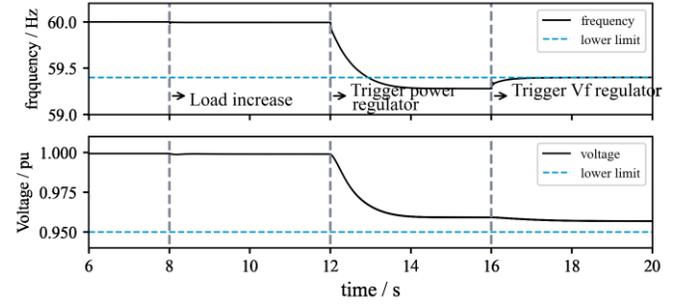

(a)

(b)

(c)

Fig. 12. Performance of the proposed control framework in Scenario 2-2: (a) dynamic inverter output; (b) static inverter output; (c) dynamic frequency and voltage

response to V-f regulation (load power decreases as V-f decrease). S1-S3 remain unchanged after triggering the V-f regulator at 16s, indicating that the power regulator and the V-f regulator work well together.

Figs. 11(b) and 12(b) show the static operating points at 6 s, 10 s, 14 s, and 18 s. After the power regulator is turned on, the circle and square operations of G1-G3 all fall on the constrained boundary.

Figs. 11(c) and 12(c) show the dynamic frequency and voltage measured at the terminal of G3. Both frequency and voltage have an obvious dip at 12 s because the total load is larger than the generation capacity. To achieve load-generation balance again, the load power decreases after the voltage and frequency dip. Scenario 2-1 and Scenario 2-2 have different V-f regulation results due to distinct load change. Scenario 2-1 has more reactive power increase and thus a larger voltage dip, while Scenario 2-2 has more active power increase and thus a larger frequency dip. From 12 s to 16 s, the voltage in Scenario 2-1 and the frequency in Scenario 2-2 violate the requirements of V-f deviation. The violations are mitigated after the trigger of the V-f regulator at 16 s without performing any load shedding.

*2) Comparison with the current limiter*

Like the comparisons in Scenario 1, the threshold of the current limiter is adaptively updated according to the real-time DER capacity, and active power priority is implemented. We discovered that triggering three current limiters simultaneously in Scenario 2 would result in instability. Hence, the current limiter of G3 is triggered 0.5 s later than those of G1 and G2. In addition, the threshold for G3 was gradually decreased. Then, the comparison results of Scenarios 2-1 and 2-2 are shown in Figs. 13 and 14, respectively. Based on the comparisons, we have the following observations:

- The current limiter causes large P-Q and V-f deviations at 8 s, while our proposed method has a smoother state transition due to the implementation of a PI controller.
- The capacity constraints of the current limiter were not accurate due to the voltage deviation. Scenario 2-1 has a large voltage dip in Fig. 13(c), and the inverter generation were inside the capacity circle at 14 s, as shown in Fig. 13(b).
- The current limiter failed to reallocate the constrained generation capacity. Hence, to recover voltage and frequency, load shedding is required after 16 s.

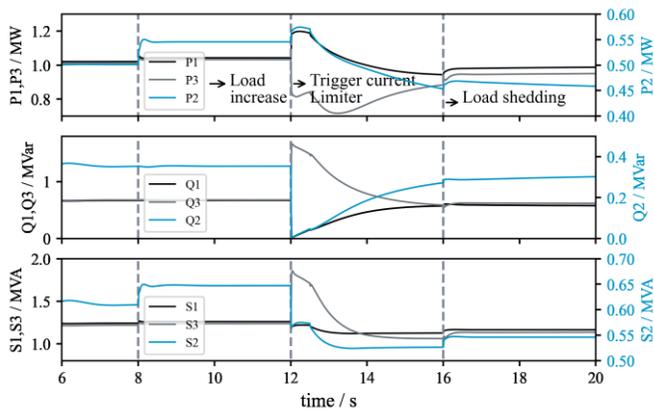

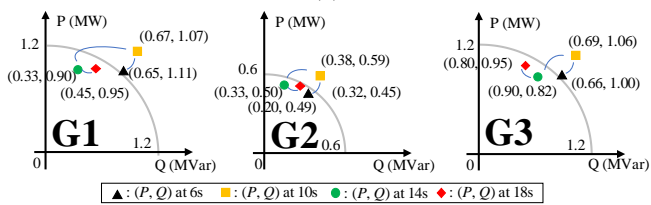

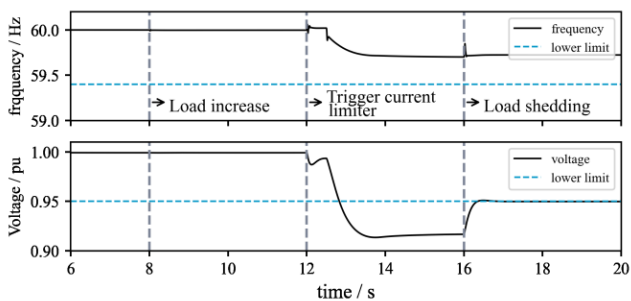

Fig. 13. Performance of current limiter in Scenario 2-1 (a) dynamic inverter output; (b) static inverter output; (c) dynamic freq. & voltage.

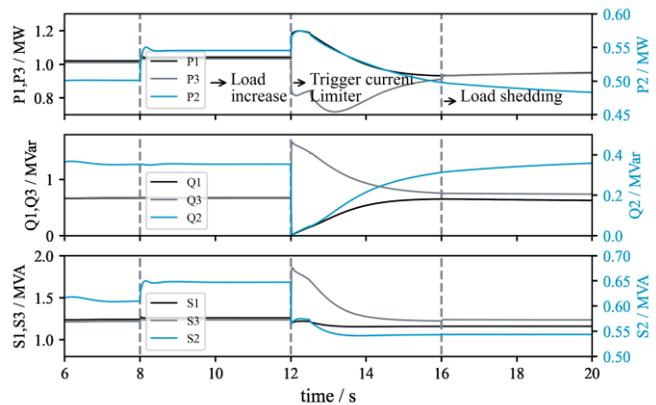

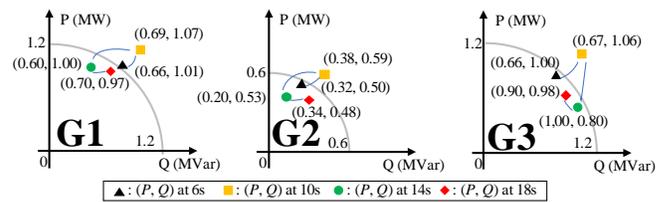

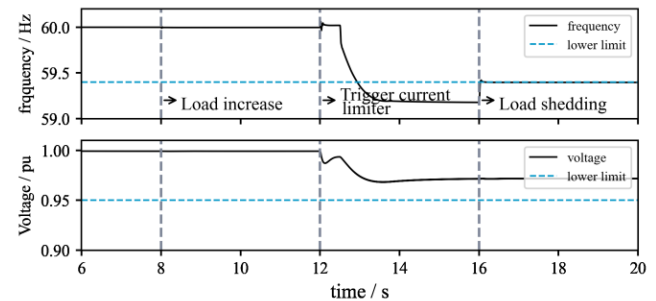

Fig. 14. Performance of current limiter in Scenario 2-2 (a) dynamic inverter output; (b) static inverter output; (c) dynamic freq. & voltage.

In Scenario 2-1, an additional counterintuitive observation has been made: the output of the inverters increased following load shedding. After load shedding, the grid voltage increased, and since the current was constrained by the limiter, the actual output increased.

In general, Scenarios 2-1 and 2-2 verify the proposed decentralized and coordinated control framework. The framework enables output control of GFM inverters and V-f improvement under the condition of constrained DER capacity and outperforms the current limiter.

## VI. CONCLUSION

This paper proposes a decentralized and coordinated V-f control framework to address the challenges brought by DER inadequacy in islanded microgrids. The control framework is composed a power regulator and a V-f regulator. The power regulator can regulate the output of GFM inverters without violating real-time reference capacities, while the V-f regulator improves the V-f deviation by leveraging the load response to voltage and frequency. The control framework is developed based on droop control and is thus purely decentralized and not dependent on costly communication interfaces. In addition, it guarantees DC voltage stability and reduces involuntary load shedding. Within the proposed control framework, three-level coordination is achieved simultaneously when DERs are inadequate, including 1). Generation and load, 2). V regulation and f regulation, and 3). P generation and Q generation.

This paper focuses on V-f control at the primary level for balanced microgrid. Future works may apply the fundamental idea to unbalanced systems and develop detailed energy management strategies that generate the real-time constrained capacity.

## VII. ACKNOWLEDGEMENT

The authors would like to thank the financial support in part from the US DOD ESTCP program under the grant number EW20-5331 to complete this research work.